\documentclass[conference]{IEEEtran}
\IEEEoverridecommandlockouts
\usepackage{listings}
\usepackage{url}
\usepackage{hyperref} 
\usepackage{cite}
\usepackage{amsmath,amssymb,amsfonts}
\usepackage{algorithmic}
\usepackage{graphicx}
\usepackage{textcomp}
\usepackage{xcolor}
\usepackage{multirow}
\usepackage{xcolor}
\usepackage[T1]{fontenc} 
\def\BibTeX{{\rm B\kern-.05em{\sc i\kern-.025em b}\kern-.08em
    T\kern-.1667em\lower.7ex\hbox{E}\kern-.125emX}}

\definecolor{codegreen}{rgb}{0,0.6,0}
\definecolor{codegray}{rgb}{0.5,0.5,0.5}
\definecolor{codepurple}{rgb}{0.58,0,0.82}
\definecolor{backcolour}{rgb}{0.95,0.95,0.92}
\definecolor{lcol}{HTML}{AF3235}

\hypersetup{
    colorlinks=true,
    allcolors=lcol,
    linkcolor=black
}

\newcommand{\websiteurl}{https://sites.google.com/view/mic-e-mouse}
\newcommand{\websitelink}[1]{\href{\websiteurl}{#1}}

\renewcommand{\paragraph}[1]{\textit{\textbf{#1}}}

\usepackage{bold-extra}
\renewcommand{\textsc}[1]{\textbf{\scshape #1}}
\newcommand{\rtextsc}[1]{{\scshape {#1}}}

\AtBeginDocument{%
  \providecommand\BibTeX{{%
    Bib\TeX}}}

\usepackage{tikz}
\usepackage{amsmath}

\usepackage{filecontents}

\usepackage{xcolor}
\usepackage{pifont}
\usepackage{url}
\usepackage{colortbl}
\usepackage{caption}
\usepackage{subcaption}
\usepackage{graphicx}
\usepackage{xurl}
\definecolor{darkgreen}{rgb}{0.0, 0.5, 0.0}

\usepackage{ulem}
\usepackage{graphicx}
\usepackage{booktabs}
\usepackage{multirow}

\lstdefinestyle{mystyle}{
    language=c,
    backgroundcolor=\color{backcolour},   
    commentstyle=\color{codegreen},
    keywordstyle=\color{magenta},
    numberstyle=\tiny\color{codegray},
    stringstyle=\color{codepurple},
    basicstyle=\ttfamily\footnotesize,
    breakatwhitespace=false,         
    breaklines=true,                 
    captionpos=b,                    
    keepspaces=true,                 
    numbers=left,                    
    numbersep=5pt,                  
    showspaces=false,                
    showstringspaces=false,
    showtabs=false,                  
    tabsize=2
}

\begin{document}

\title{\textit{Invisible Ears at Your Fingertips:}\\
Acoustic Eavesdropping via Mouse Sensors
}

\author{\IEEEauthorblockN{Mohamad Habib Fakih,
Rahul Dharmaji, Youssef Mahmoud, Halima Bouzidi, Mohammad Abdullah Al Faruque}
Dept. of Electrical Engineering and Computer Science, University of California, Irvine, CA, USA \\
\{mhfakih, rdharmaj, yhmahmou, hbouzidi, alfaruqu\}@uci.edu
}

\maketitle

\begin{abstract}
Modern optical mouse sensors, with their advanced precision and high responsiveness, possess an often overlooked vulnerability: they can be exploited for side-channel attacks. This paper introduces Mic-E-Mouse, the first-ever side-channel attack that targets high-performance optical mouse sensors to covertly eavesdrop on users. We demonstrate that audio signals can induce subtle surface vibrations detectable by a mouse's optical sensor. Remarkably, user-space software on popular operating systems can collect and broadcast this sensitive side channel, granting attackers access to raw mouse data without requiring direct system-level permissions. Initially, the vibration signals extracted from mouse data are of poor quality due to non-uniform sampling, a non-linear frequency response, and significant quantization. To overcome these limitations, Mic-E-Mouse employs a sophisticated end-to-end data filtering pipeline that combines Wiener filtering, resampling corrections, and an innovative encoder-only spectrogram neural filtering technique. We evaluate the attack's efficacy across diverse conditions, including speaking volume, mouse polling rate and DPI, surface materials, speaker languages, and environmental noise. In controlled environments, Mic-E-Mouse improves the signal-to-noise ratio (SNR) by up to +19 dB for speech reconstruction. Furthermore, our results demonstrate a speech recognition accuracy of roughly 42\% to 61\% on the AudioMNIST and VCTK datasets. All our code and datasets are publicly accessible on \websitelink{Mic-E-Mouse website}\footnote{\websitelink{https://sites.google.com/view/mic-e-mouse}}.
\end{abstract}

\section{Introduction}

The proliferation of low-cost, high-fidelity sensors in consumer devices has greatly improved user experience in common computing tasks. From lower response times to more adaptive workflows, these devices have increased productivity while remaining affordable to the average consumer. The lion's share of these improvements is found in the category of user input devices, including styli~\cite{ngstyli, kodakstyli}, mice~\cite{faresmouse,xiangmouse}, and monitors \cite{amirmonitor,marcumonitor}. More specifically, improvements in mice sensor technologies have allowed commercial offerings to operate with a sample rate of 4KHz \cite{darmosharkm3}, with a growing selection of products that also support 8KHz \cite{razerviper}.

\begin{figure}[tp]
    \centering
    \includegraphics[width=0.45\textwidth]{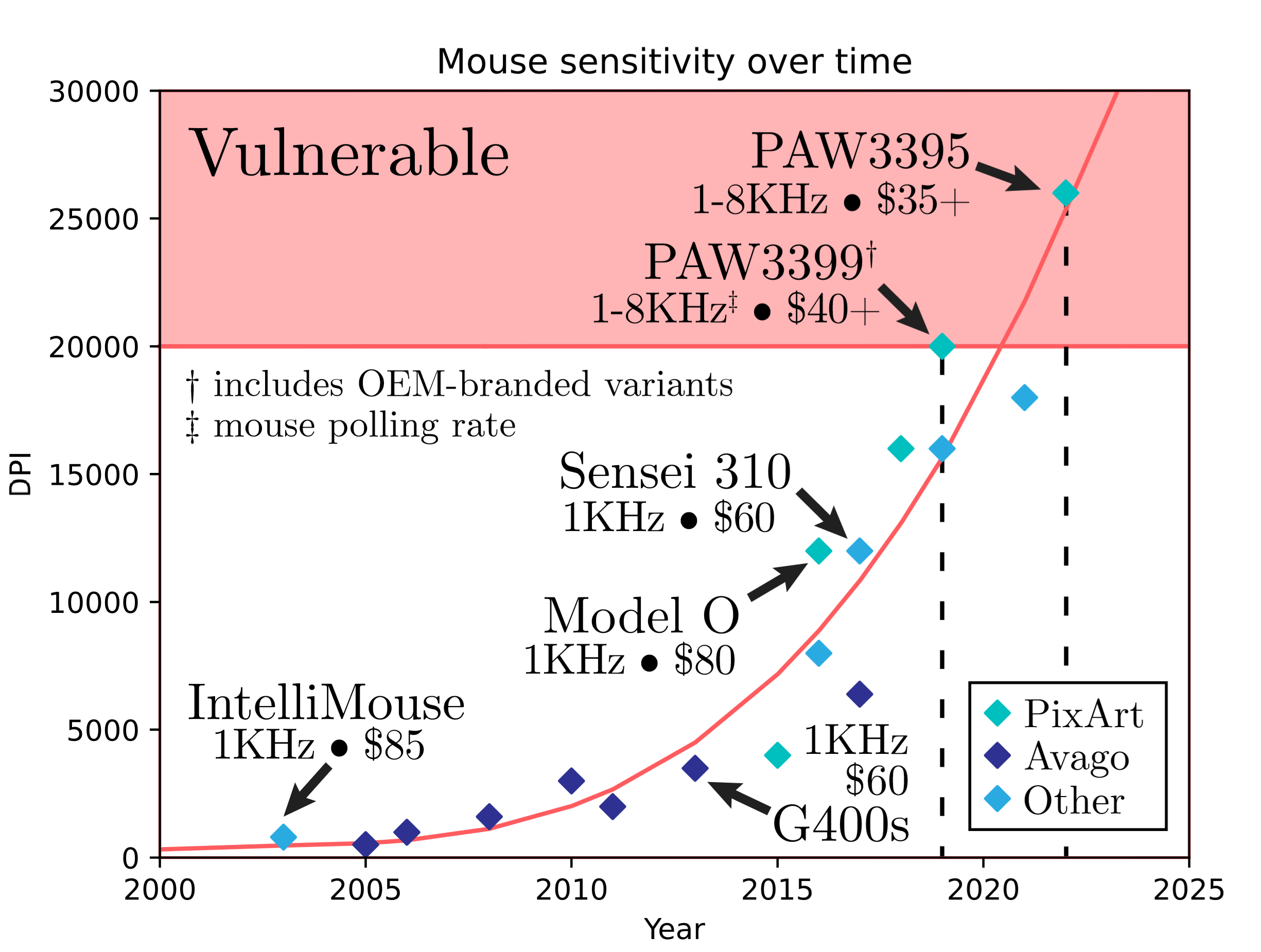}
    \caption{Computer mice optical sensor fidelity trends over time. The red-shaded region indicates vulnerable sensors featuring high resolution measured in DPI (\textit{Dots-per-inch}).}
    \label{fig:mice_over_time}
    \vspace{-0.5cm}
\end{figure}

Consumer-grade mice with high-fidelity sensors are already available for under 50 U.S. Dollars \cite{darmosharkm3}. As improvements in process technology and sensor development continue, it is reasonable to expect further price declines, similar to the trend shown in Figure \ref{fig:mice_over_time}. Furthermore, mouse sensors' resolution and tracking accuracy also follow the same pattern, with steady improvements each year. Ultimately, these developments entail an increased usage of \textit{vulnerable mice} by consumers, companies, and government entities, expanding the attack surface of potential vulnerabilities in these advanced sensor technologies.

The rise in \textit{work-from-home} policies has led to the widespread adoption of new technologies and practices, making it more difficult for employers and government institutions to control the physical operating environments of their workforce.
While these arrangements often boost employee sentiment and productivity~\cite{klinewfh}, the security implications of \textit{work-from-home} policies are still being understood~\cite{borkewfh}. Specifically, attacks exploiting personal peripherals on work computers, such as keyboards~\cite{zhukeyboard, asonovkeyboard}, microphones~\cite{genkinlend}, styli~\cite{farrukhpencil,liumaghacker}, earphones~\cite{liecho}, mechanical hard drives~\cite{kwonghdd}, and even USB devices~\cite{dumitruusb}, have become increasingly common. Even in relatively secure office environments, the threat posed by these exploits is still significant, especially for unknown or poorly understood attack vectors.

\begin{figure}[tp]
    \centering
    \includegraphics[width=0.48\textwidth]{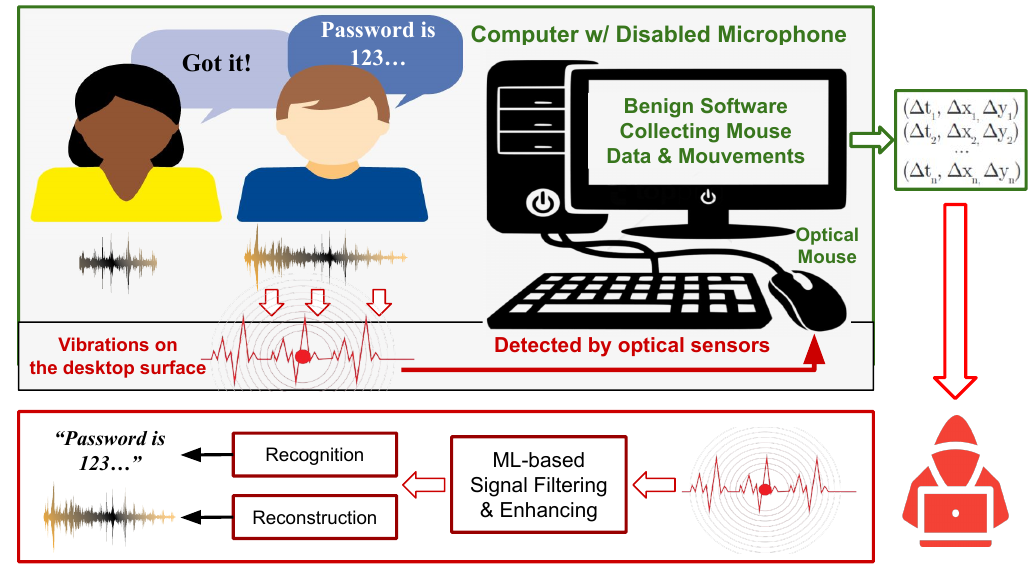}
    \caption{Overview of the Mic-E-Mouse pipeline: A victim's confidential speech is captured by benign or compromised software using surface vibrations detected by the computer mouse. The collected data is sent to the adversary's server for processing and filtering with machine learning methods to enhance the quality of the recovered audio signal.
    }
    \vspace{-0.5cm}
    \label{fig:pipeline_overview}
\end{figure}

We posit that the seemingly innocuous computer mouse is the source of yet another vulnerability. Importantly, we claim recent advancements in mouse sensor resolution can be sufficient to enable a side-channel attack capable of extracting user speech. 
Through our Mic-E-Mouse pipeline, vibrations detected by the mouse on the victim user's desk are transformed into comprehensive audio, allowing an attacker to eavesdrop on confidential conversations. 
This process is stealthy since the vibrations signals collection is invisible to the victim user and does not require high privileges on the attacker's side. Potential adversaries can collect \textbf{user-space} mouse signals and remotely use the Mic-E-Mouse pipeline to convert raw data packets into audio. Figure \ref{fig:pipeline_overview} provides an overview of our proposed attack pipeline.

We summarize our key contributions as follows:

\begin{itemize}
    
    \item To the best of our knowledge, our work is the first to exploit computer mouse sensors as a means to capture confidential data (i.e., user speech) from environmental vibrations, even in highly secure environments without any additional conventional recording devices.

    \item We introduce the \textbf{MIC}rophone-\textbf{E}mulating-\textbf{MOUSE}, or "Mic-E-Mouse"\footnote{The name "Mic-E-Mouse" was inspired from a certain fictional mouse with big ears.}, a pipeline consisting of successive signal processing and Machine Learning stages to overcome the challenges involved in speech extraction from extracted mouse data, ultimately enabling intelligible reconstruction of user speech.

    \item We present a detailed study on the motivations, methodologies, and efficacy of the \textbf{Mic-E-Mouse} pipeline through comprehensive testing with high-performance mice and optical sensors while considering the effects of varying surface types, polling rates, distance, and mouse sensitivity configurations.
\end{itemize}

\section{Background}
\label{bg}

This section provides the required background knowledge, including optical computer mice technology, operating-system-level input drivers, and machine learning techniques for speech signal processing and enhancement.

\subsection{Optical Computer Mice}
\label{sec:MouseInternals}
Modern optical mice employ various methods to provide precise movement tracking under different sensitivity settings. Over the last two decades, optical mice leveraging a high-performance CMOS camera with an onboard \textit{Digital Signal Processor} (DSP) have become the preferred design choice~\cite{palacinoptical}. Generally, optical sensors enhance reliability and fidelity through the use of self-illumination, typically from an independent diode or an integrated laser. By taking thousands of snapshots of the illuminated surface under the mouse, the DSP can then compare each successive image in order to determine the direction of movement. The rate at which this process happens is determined by the sensor's frame rate, measured in \textit{Frames Per Second} (FPS). Each frame is processed via an on-chip correlation algorithm to provide a 2-dimensional displacement to the host computer~\cite{gordondsp}. The described process can be broken down into two key elements: (\emph{i}) the imaging sensors and (\emph{ii}) the image processing and movement detection algorithm.

\begin{figure}[ht!]
    \centering
    \includegraphics[width=0.8\columnwidth]{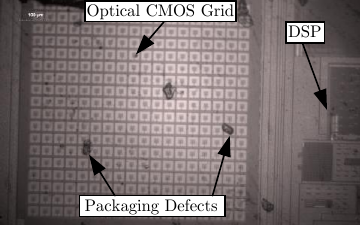}
    \caption{Imaging of the CMOS sensor grid in the PMW3552 chip from a Logitech mouse}
    \vspace{-0.3cm}
    \label{fig:bg}
\end{figure}

\noindent \textbf{Imaging Sensors.} Rather than relying on expensive \textit{Charge-Coupled Device} (CCD) sensors, the sensor in an optical mouse is typically a \textit{Complementary Metal-Oxide-Semiconductor} (CMOS) image sensor collecting up to 30x30 pixels' worth of data per frame, where each pixel represents the intensity of the reflected light at that point \cite{cmossensor}. This basic mini-camera is a critical component for implementing speckle-pattern detection. Some sensor models, such as the PixArt PMW3552, capture data using an 18×18 pixel grid, while others can record up to 30×30 pixels, depending on the manufacturer’s specifications. For visualization purposes, we destructively studied a PixArt PMW3552 sensor in our institutional lab, and the resulting scan is shown in Figure~\ref{fig:bg}. This sensor features an 18×18 CMOS pixel grid and is designed to interface directly via USB. Speckle patterns are random, granular intensity patterns produced when coherent light, such as laser light, is scattered by a rough surface \cite{stachowiakspeckle}. When an optical mouse is moved across a surface, the speckle pattern on the surface changes smoothly and reliably. The CMOS sensor captures these changes in the speckle pattern frame by frame and processes them to detect movement\cite{cmosspeckle}. These movement detection algorithms allow for the translation of data into corresponding coordinate deltas.

\noindent \textbf{Image Processing and Movement Detection.} Movement detection involves analyzing the changes between successive images. This can be modeled as:
\begin{equation}
\vec{\Delta p} = f(\vec{I}_t, \vec{I}_{t-1})
\end{equation}

where $\vec{\Delta p}$ is the displacement vector of the mouse, $f$ is the image processing function, $\vec{I}_t$ is the image at time $t$, and $\vec{I}_{t-1}$ is the image at time $t-1$.
The function $f$ often involves correlating sections of the two images to estimate movement. This correlation can be represented as:
\begin{align}
C(\vec{x}) &= \sum_{\vec{u} \in W} \vec{I}_t(\vec{u}) \cdot \vec{I}_{t-1}(\vec{u} - \vec{x}) \\
\vec{\Delta p} &= \underset{\vec{x}}{\text{argmax}} \: C(\vec{x}) \notag
\end{align}
where $C(\vec{x})$ is the correlation for displacement $\vec{x}$, and $W$ is the window of pixels in the image being analyzed. 

The use of the \textit{Fast Fourier Transform} (FFT) significantly enhances the efficiency of calculating the correlation function~\cite{fft}. In the case of image correlation, the FFT is applied to both images ($\vec{I}_t$ and $\vec{I}_{t-1}$) to transform them into the frequency domain. This transformation simplifies the calculation of the cross-correlation, which is computationally less intensive in the frequency domain~\cite{ft}. The equation for cross-correlation in the frequency domain can be expressed as:
\begin{equation}
C(\vec{x}) = \mathcal{F}^{-1}\left\{\mathcal{F}\left\{\vec{I}_t\right\} \circ \mathcal{F}\left\{\vec{I}_{t-1}\right\}^*\right\}
\end{equation}

where $\mathcal{F}$ and $\mathcal{F}^{-1}$ represent the Fourier Transform and its inverse, respectively, and $^*$ denotes the complex conjugate.

The displacement vector $\vec{\Delta p}$, once obtained, is scaled by the DPI setting of the input device to ensure accurate correspondence between physical movement and on-screen cursor movement~\cite{boudmouse, pangmouse}. This vector is further refined through Kalman filtering, a recursive algorithm enhancing motion tracking by reducing noise in the trajectory of the displacement vector~\cite{kalmanfilter, likalman}. Finally, this scaled and smoothed vector undergoes quantization in the mouse's firmware, converting it from a continuous to a discrete form for compatibility with digital systems, thereby maintaining cursor movement accuracy and fluidity, with high-performance mice featuring custom enhancements for input quality.

\begin{figure}
    \centering
    \includegraphics[width=\linewidth, trim=0.7px 0.7px 0.7px 0.7px, clip]{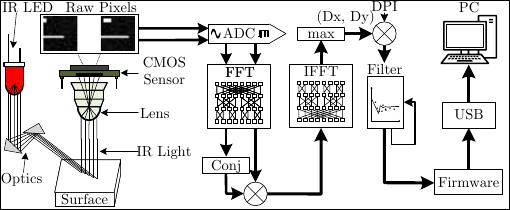}
    \caption{Overview of the internal systems of a mouse.}
    \vspace{-0.5cm}
    \label{fig:MouseInternals}
\end{figure}

\subsection{Operating System Input Handling}
\label{os}
Modern operating systems, such as those using the Linux kernel, as well as Windows and macOS, handle peripheral device input through interrupt-based drivers, allowing them to efficiently manage and respond to user interactions with input devices such as the computer mouse. Additionally, these operating systems separate access to input devices into two distinct areas: kernel space and user space.

\subsubsection{Kernel Space System Calls} \
Kernel space is a privileged area of the system's memory where the core components of the operating system such as the kernel and device drivers- operate. Logging the mouse movements in kernel space is possible through certain OS-dependent endpoints.

\noindent \textbf{Linux kernel} provides a method of obtaining input data through kernel modules such as \verb|hid_logitech_hidpp|. This driver is responsible for collecting mouse events and relaying them to the kernel's input subsystem. The Linux kernel uses the following  endpoint to provide access to mouse data \verb|/dev/input/mouseX|. Normally, direct access to this endpoint requires elevated privileges. 
The Linux kernel also provides a mechanism in order to uniquely identify mouse input devices. Consumer mice have a predetermined name set by the manufacturer, which can be used to obtain the correct input endpoint. Given a target mouse with a vulnerable sensor (e.g. the Razer Viper 8KHz used in our experimental setup detailed in Table \ref{table:usedmice}), we can query the \verb|/proc/bus/input/devices| input for the corresponding name. By doing so, we can automatically obtain the input endpoint required for an exploit.

\noindent \textbf{Windows} operating system also provides a method to retrieve low-level mouse events. By using the \verb|LowLevelMouseProc| function described in the Windows API, a program can obtain displacement information directly from the operating system \cite{winapimouse}. The function yields a data struct of the \verb|MSLLHOOKSTRUCT| type, containing both the displacement and button information. 

\noindent \textbf{MacOS} operating system provides a method to capture low-level mouse events through the Quartz Event Services framework. By using the \verb|CGEventTapCreate| function, developers can create an event tap that monitors or intercepts mouse events such as clicks, movements, and scrolls at the system level. 

\subsubsection{User Space Applications}
Modern operating systems provide a user-space access to input devices. An adversary readily get the data logged by any peripheral (e.g., computer mouse) without necessitating elevated permissions. Many software libraries already provide such access points. Graphical frameworks like \verb|Qt|\cite{Qt}, \verb|GTK|\cite{GTK}, and \verb|SDL|\cite{SDL} enable real-time mouse data access from unprivileged user-space applications. Similarly, game engines like Unity\cite{unity}, PyGame~\cite{pygame}, and Unreal~\cite{unreal} offer extensive mouse tracking support and built-in networking capabilities, further facilitating an attacker's efforts.

\section{Threat Model}
\label{threatmodel}

\begin{figure*}[ht!]
    \centering
    \includegraphics[width=\linewidth]{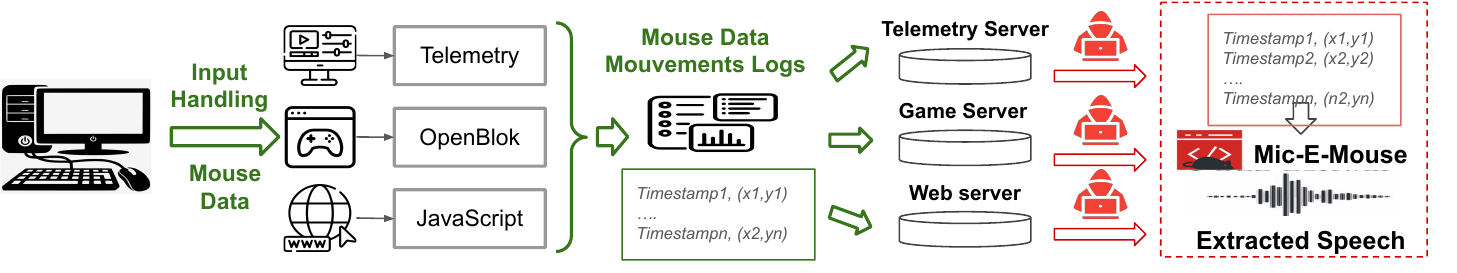}
    \caption{An overview of a practical attack scenario following the proposed Mic-E-Mouse pipeline with different vulnerability exploits, including, graphical application, open-source games, and web browser. Green and Red arrows depict authorized and unauthorized access, respectively.
    }
    \label{fig:example_exploit}
    \vspace{-0.3cm}
\end{figure*}

\noindent \textbf{Attack Scenario and Goals}. We consider a practical attack scenario in which the victim is engaged in a sensitive conversation within a room equipped with a desktop computer without microphone or any other voice recording equipment. The attacker can be an individual who aims to eavesdrop on the confidential conversation from outside the room. The attacker can exploit the victim's computer mouse to recover speech signals from the surface vibrations detected by the highly-sensitive optical mouse sensors. The recovered speech could be exploited by the eavesdropper for various malicious purposes (e.g., intelligence-gathering or blackmailing). 

\noindent \textbf{Attack Assumptions}: We assume that the victim is in a controlled environment, such as a private office or home workspace, where they use a desktop computer equipped with a commercially available optical mouse. The mouse is assumed to have high sensitivity (e.g., a sensor of at least 20,000 DPI) and a polling rate of at least 8 KHz, specifications commonly found in mid-range to high-performance input devices (See Appendix \ref{sec:additionalmice}). The work surface is assumed to allow vibrations to propagate to the mouse sensor (e.g., a thickness of no more than \textbf{3cm}), which is typical for wooden or laminate desks with standard thickness. 
The victim's computer system is assumed to run an application or process capable of logging pointer movements, such as background services, custom software, or system utilities, which may or may not require elevated privileges. This logging capability is presumed to exist either as a legitimate feature or via unauthorized exploitation discussed in Section \ref{sec:exploit}.
During the attack scenario, the victim is assumed to engage in activities involving sensitive spoken information. While the victim may intermittently use the mouse, the attack leverages instances where the pointer is stationary or experiences minimal motion. The conversation volume is assumed to fall within a range of 60–80dB, which covers typical office or home discussions \cite{nassilamp, LidarPhone, gyrophone}. The attacker is assumed to have access to mouse movement data via software-level logging, remote exploits, or physical access to the machine. To strengthen the practicality of the threat model, we discuss in Section \ref{sec:exploit} multiple exploits, including but not limited to compromised applications, unauthorized monitoring software, or vulnerabilities in the input data handling pipeline.

\subsection{Vulnerability Exploitation Methods}
\label{sec:exploit}
In this section, we discuss multiple exploit methods that can be leveraged by the attacker to access the history of mouse movement logs. Specifically, we explore the following vulnerability exploitation methods:

\subsubsection{Vulnerability Exploit in Graphical Applications}:
Software that naturally gathers high-frequency mouse movement data as part of its normal operation can be exploited for the Mic-e-Mouse attack. Images/video editing applications are potential candidates due to their reliance on precise input handling. Examples of such applications include Blender \cite{swblender}, Kdenlive \cite{swkdenlive}, and Krita \cite{swkrita}, which utilize mouse data for features like fine-grained control and graphical interface interactions. Many applications with graphical interfaces collect telemetry data, which may include mouse movement logs for diagnostic purposes. While privacy policies typically regulate such data collection, telemetry practices vary widely. In some cases, raw input data could be transmitted to external servers for processing or analysis. A practical attack could involve compromising a telemetry server to gain access to raw mouse movement data. This data, if sufficiently high in resolution, could be analyzed to infer sensitive information, such as speech patterns. A detailed example of this attack and exploit are depicted in Figure \ref{fig:example_exploit}.

\subsubsection{Vulnerability Exploit in Open-source Games}:
Anti-cheat systems in gaming, such as Vanguard \cite{vanguard}, often analyze user input, including mouse movements, to detect tampering or unfair gameplay. In some cases, this analysis involves transmitting input data to remote servers managed by the game developer. While not all anti-cheat systems store raw mouse logs, those that do may create a potential attack vector. However, testing such scenarios is challenging due to the proprietary nature of most closed-source games.
For our proof-of-concept attack, we utilize \textit{OpenBlok}, an open-source version of the classic \textit{Tetris} video game \cite{openblok}. Open-source software provides a transparent environment for controlled research and experimentation. Our exploit involves applying a single patch to the game's source code before compilation. This patch introduces a background thread that captures mouse input data during gameplay and transmits it to a remote server using integrated telemetry. The exploit does not require physical access to the victim's machine, showcasing its potential applicability in real-world scenarios.

\subsubsection{Vulnerability Exploit in Web-Browser}:
In web browsers, JavaScript can be exploited to extract high-frequency mouse data. Modern web browsers generally limit input event rates to the refresh rate of the display, which is typically 60Hz for standard monitors. This limitation ensures consistency and minimizes excessive resource usage. While high-refresh-rate monitors (e.g., 120Hz, 144Hz) allow higher event rates, these remain insufficient for the Mic-E-Mouse vulnerability, which requires significantly higher frequencies.
From our preliminarily analysis, we observed that activating the developer console (\textit{F12}) in Chromium-based browsers temporarily increased the event rate to 1 KHz. This behavior appears to be an undocumented feature intended for debugging purposes, as it aligns with the requirements of high-performance development workflows. 
Currently, no mainstream web browser natively supports the high-frequency data access required to exploit the Mic-E-Mouse vulnerability. This finding demonstrated the importance of existing input rate limitations in preventing such attacks. However, future updates or changes to browser input handling—whether intentional or accidental—could reintroduce such vulnerabilities. 

\section{The Microphone-Mouse Analogy}
\label{sec:mousemicrophone}

In this section, we draw parallels between a computer mouse on a surface and a microphone with its diaphragm, demonstrating how mouse sensors can function as microphones. We explore how a mouse sensor detects transverse and longitudinal waves and the key characteristics that enable this side-channel vulnerability.

\subsection{Microphone Mechanics}

Microphones convert sound waves into electrical signals through diaphragm vibrations caused by air pressure changes. The diaphragm's movement is then transduced into electrical currents via different mechanisms depending on the microphone type~\cite{hoppermic}. The left side of Figure~\ref{fig:micmouse} depicts the internals of a basic microphone. Typically, dynamic microphones generate signals by moving a coil within a magnetic field, whereas condenser microphones alter capacitance by using a vibrating diaphragm as one plate of a capacitor.

\subsubsection{How Mice Mimic Microphones}

\begin{figure}[!ht]
    \centering
    \includegraphics[width=0.8\linewidth]{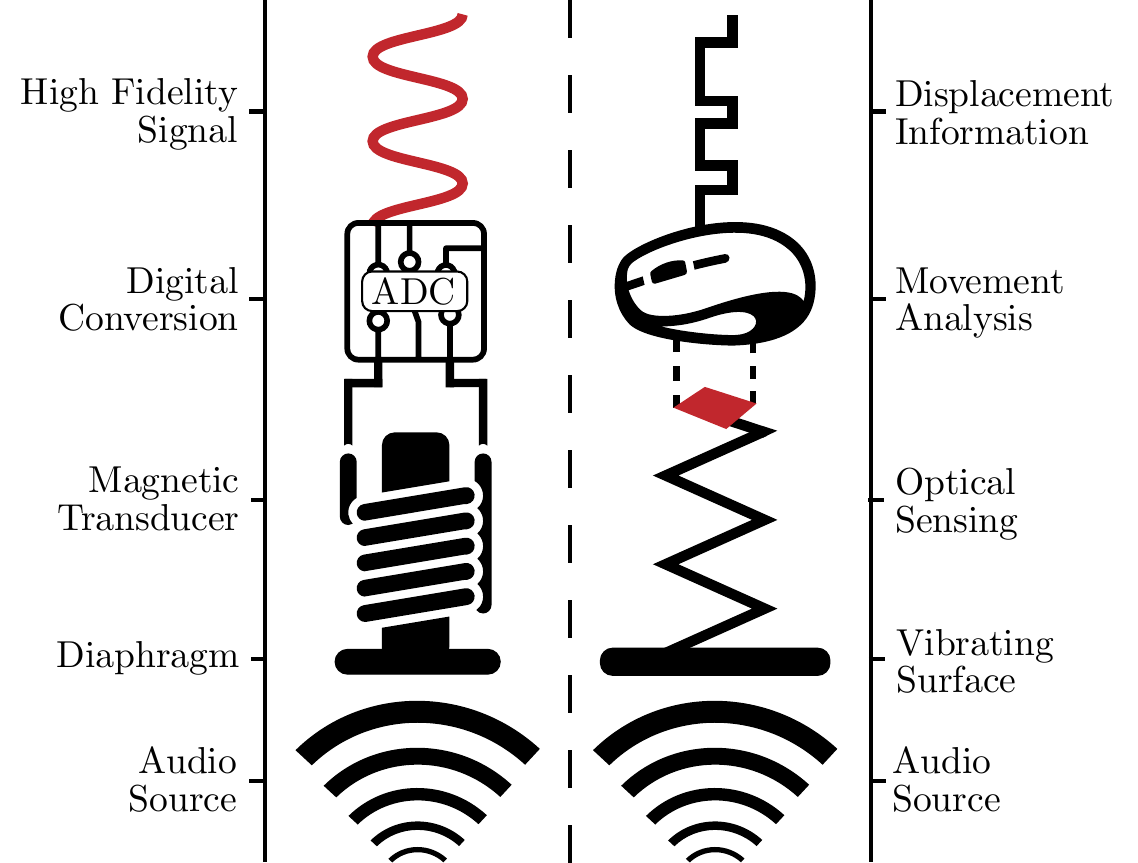}
    \caption{The internals of a high-performance mouse can be viewed as similar to those of a high-fidelity microphone.
    }
    \label{fig:micmouse}
    \vspace{-0.1cm}
\end{figure}

Computer mice are not originally designed to capture sound, but their sensors can detect surface vibrations similar to how a microphone’s diaphragm captures sound waves~\cite{gordondsp}. When a mouse is stationary, these vibrations can resemble ambient noise picked up by a microphone. An optical mouse's laser or LED illuminates the surface while the sensor, functioning like a camera, captures thousands of images per second. The mouse's built-in processor interprets changes between these images to detect movement, which may inadvertently contain sound-induced signals.
The unintended microphone-like behavior occurs when the surface vibrations are within the detectable range of the mouse's sensor. These vibrations cause minute movements of the mouse relative to the surface, which the sensor interprets as movement. 
With the right signal processing, these signals can be converted back into sound waves, effectively turning the mouse into a rudimentary microphone. Although mice can track these minute vibrations under controlled conditions, real-world usage and surface irregularities can dramatically reduce viability. We thus emphasize that these results represent an upper bound under idealized setups. 
We show the key similarities between a microphone and the proposed Mic-E-Mouse makeshift microphone in Figure \ref{fig:micmouse}. 
Although conceptually similar to a microphone, the mouse sensor is not designed to respond to air pressure changes. Instead, it passively measures surface displacement, requiring far stronger coupling between acoustic vibrations and the work surface for effective signal capture.

\subsubsection{Decoding Vibrations}

\begin{figure}[!ht]
    \centering
    \includegraphics[trim=1px 1px 1px 1px, clip]{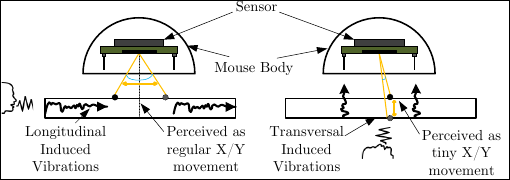}
    \caption{Model of how sound waves propagate on the surface.}
    \label{fig:micasmouse}
    \vspace{-0.3cm}
\end{figure}

The process by which a computer mouse detects vibrations and converts them into signals interpretable as sound involves two key components. (\emph{i}) \textit{Longitudinal vibrations}, which move parallel to the surface, cause the most significant movements of the mouse sensor and are therefore more easily interpreted as traditional cursor movements. (\emph{ii}) \textit{Transverse vibrations}, move perpendicular to the surface, and result in much subtler signals. These are usually ignored during normal mouse operation but can contain the range of frequencies necessary for reconstructing sound.
For a mouse to function as a microphone, its sensor must be sensitive enough to detect \textit{transverse vibrations}. The DPI (Dots Per Inch), which measures the sensor's resolution over a given distance, is a crucial metric to determine the required level of sensitivity. High-DPI mice (10,000+ DPI) are more likely to capture subtle vibrations similar to those detected by a microphone's diaphragm. This behavior is detailed in Fig. \ref{fig:micasmouse}.

\begin{figure*}
    \centering
    \includegraphics[width=\textwidth]{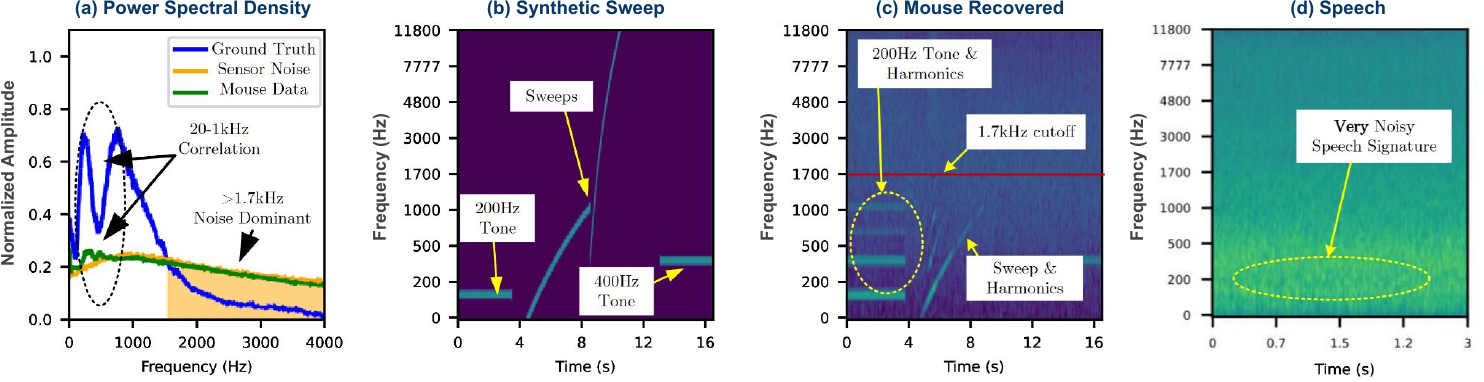}
    \caption{Signal composition for different setups: \textbf{(a)} Frequency composition of the ground truth data, sensor noise from the mouse, and the data as collected with the mouse \textbf{(b)} Spectrogram of the synthetic sweep signal we generate \textbf{(c)} spectrogram of the sweep signal as collected by the mouse. \textbf{(d)} Spectrogram of an unfiltered speech signal as collected by the mouse.}
    \label{fig:noise}
    \vspace{-0.2cm}
\end{figure*}

For a mouse with a given DPI and polling rate, the smallest detectable movement is determined by its resolution. A mouse with a DPI of $D$ can detect movements as small as $1/D$ inches. The polling rate $S$, measured in Hz, is the rate at which the mouse sensor updates the host PC. For transverse waves on a surface with frequency $f$, the mouse sensor will detect these vibrations if their frequency is below the Nyquist frequency, which is half the polling rate. According to the Nyquist-Shannon sampling theorem \cite{shannonfs}, this is the maximum frequency that can be accurately sampled without aliasing. For a wave with frequency $f$, wavelength $\lambda$ on the surface can be calculated if we know the speed of the wave on the surface $\nu$:
\begin{equation}
    \lambda = \frac{\nu}{f}
\end{equation}

\noindent The DPI must be high enough so that the smallest detectable movement is less than half the wavelength of the wave (We convert D into millimeters by since $1in = 25.4mm$):
\begin{align}
    \frac{25.4}{D} &< \frac{\lambda}{2} \notag \\
    \frac{f}{D} &< \frac{\nu}{50.8}
\end{align}

\noindent If DPI and polling rate are high enough to satisfy this condition, the mouse sensor can detect the wave and thus capture the transversal waves. Assuming a mid-range sound velocity in wood (e.g., 3000±500 m/s), 
we demonstrate that our calculations hold under common desk conditions, and that we can expect to recover sounds from the mouse given $f=8kHz$ and $D=20000$.

\subsection{Preliminary Study}

We perform a preliminary study to confirm that the mouse can detect frequencies within the range of human speech---typically between 200 and 2 KHz \cite{bishopfreq, speechbanana}.
We generate a step-wise sweep from 0 to 16 kHz to observe any aliased components at frequencies above the Nyquist limit (4 kHz), thus testing how the mouse sensor responds to out-of-band signals. The sweep is generated using a synthetic \textit{tone and sweep} audio sample which consists of four consecutive segments: (\emph{i}) an initial 200Hz tone lasting 5 seconds, (\emph{ii}) a linear sweep from 0 to 1 KHz over 4 seconds, (\emph{iii}) a linear sweep from 0 to 16 KHz over 4 seconds, and (\emph{iv}) a 400Hz tone lasting 5 seconds. The start and end tones serve as a synchronization token since we observed a strong response during our initial experiments. Additionally, we cross-reference the spectral distribution of the logged mouse signals, the true data signals, and a collection of background noise in order to establish a statistical model of the noise profile for further processing during the signal processing stage of our pipeline. Figure~\ref{fig:noise} presents: (a) spectral densities of ground truth speech, ambient sensor noise, and speech collected from the mouse; (b) the ground truth synthetic sweep; (c) the sweep collected from the mouse; and (d) a speech signal collected from the mouse. In Figure \ref{fig:noise}(a), the mouse-captured signal exhibits two prominent peaks that align in frequency with the ground truth but differ in amplitude, indicating partial resonance unique to the sensor. Meanwhile, panel (c) highlights aliased harmonics that provide further evidence of how effectively (or poorly) the mouse replicates subtle audio content. The mouse sensor shows a \textbf{non-Gaussian} noise profile with a power peak around 1 kHz. Figures~\ref{fig:noise}b and~\ref{fig:noise}c demonstrate a 'tone and sweep' test, with the mouse-collected data mirroring the ground truth pattern, confirming the presence of a side-channel. In Figure~\ref{fig:noise}d, speech signals appear faint but distinct below 500 Hz, indicating the need for classical and neural filtering techniques.

\section{Mic-E-Mouse Attack Design}
\label{setup}

In this section, we detail and justify the design choices that enable the Mic-E-Mouse pipeline to recover high-quality audio signals from a computer mouse. We discuss targeted mouse sensors, the methodology for collecting vibration data, the preprocessing steps to reconstruct original audio signals, and the ML techniques used to enhance waveform reconstruction.

\subsection{Targeted Mouse Sensors}

We focus on two state-of-the-art mice sensors currently available in consumer-grade mice -- \rtextsc{Paw3395} and \rtextsc{Paw3399} -- both produced by the manufacturer PixArt Imaging Inc. \cite{paw3395ds,paw3399ds}. These sensors are in the Razer Viper 8KHz \cite{razerviper} and Darmoshark M3-4KHz \cite{darmosharkm3}, respectively. An overview of the mice sensors technology used in our testing methodology is provided in Table \ref{table:usedmice}. We also list similar mice that are vulnerable to the same class of exploit as the Razer and Darmoshark mice in Appendix \ref{sec:additionalmice}. These additional mice share at least one of the following attributes: (\emph{i}) a \rtextsc{Paw3395} or \rtextsc{Paw3399} sensor or (\emph{ii}) a poll rate at or above 4KHz. 

\begin{table}[ht!]
\centering
\caption{Targeted mouse sensors' capabilities in terms of DPI (\textit{Dots-per-Inch}), IPS (\textit{Inch-per-Second}), and Polling Rate.}
\resizebox{\linewidth}{!}{%
    \begin{tabular}{r|cccc}
    \toprule
    \rtextsc{Mouse} & \rtextsc{Sensor} & DPI & IPS& \rtextsc{POLL RATE} \\
    \midrule
    Razer Viper 8KHz & \rtextsc{Paw3399} & 20,000 & 650 & 8KHz \\
    Darmoshark M3 & \rtextsc{Paw3395} & 26,000 & 650 & 4KHz \\
    AtomPalm Hydrogen & \rtextsc{Paw3360} & 12,000 & 250 & 8kHz \\
    Redragon M994 & \rtextsc{Paw3395} & 26,000 & 650 & 1kHz \\
    \bottomrule
    \end{tabular}%
    }%
\vspace{-0.3cm}
\label{table:usedmice}
\end{table}

\subsection{Mouse Data Collection Setup}

\begin{figure}[tp]
    \centering
    \includegraphics[width=.4\textwidth, trim=1px 1px 1px 1px, clip]{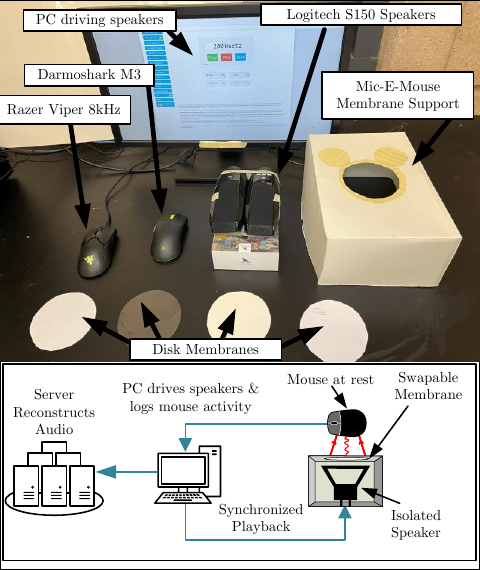}
    \caption{The Experimental setup used to collect the mouse data signals used to train and evaluate the models involved in the Mic-E-Mouse pipeline.
    }
    \label{fig:setup}
    \vspace{-0.5cm}
\end{figure}

We build a reliable data collection setup to capture mouse movements induced by playback sound at controlled audio levels using a speaker. The setup, shown in Figure~\ref{fig:setup}, includes a sound-isolation box containing a wired '\textit{Logitech S-150}' loudspeaker. Above the loudspeaker, an interchangeable membrane is centered on a 4'' ($\sim$10cm) cutout, serving to transmit vibrations to a high-performance mouse placed on its surface. The mice, listed in Table~\ref{table:usedmice}, are configured to use their highest polling rate and DPI settings to maximize sensitivity.

To collect data, each audio sample is played through the loudspeaker while the mouse's reported movement is simultaneously logged by the host computer. We used a calibrated SPL meter at the membrane surface to confirm an 80 dB SPL ±1 dB across a range of test frequencies, ensuring consistent amplitude for all initial experiments. In later experiments (Sec \ref{sec:conditions}), the amplitude was adjusted from 80 dB to explore the effects of varying sound pressure levels on the system.

In our test environment, the host computer runs Linux (kernel 6.6.3) and it collects mouse movement data via the method discussed in Section~\ref{os}. The sequence of data packets is written to a CSV file, where each collected sample contains two entries:
\begin{enumerate}
    \item A timestamp $\Delta t$, representing the time in microseconds since the last query was processed.
    \item The directional movement $\Vec{\Delta_P} \in \mathbb{R} ^ 2 = (\Delta X, \Delta Y)$, representing the discretized X and Y movement components detected by the mouse during the time since last query.
\end{enumerate}

Our data collection program processes data at up to 8kHz, which is crucial for minimizing latency as the script manages time-keeping to compute $\Delta t$.

We note that we are also able to use user-space mouse collection routines, such as those provided by the X11 display server in order to collect mouse data without requiring elevated privileges. This allows for our exploit to be embedded in applications without asking for \verb|sudo| access. Then, we can retrieve mouse events using third-party software libraries such as \verb|Qt|, \verb|GTK|, or \verb|SDL|. These libraries are commonplace in many user-facing applications, which allows our exploit to safely depend on them without raising suspicion.

\subsection{Mouse Data Preprocessing}
\label{subsec:process}
The output of the data collection program is a CSV file with only timestamp and positional movement information, which is far from an adequate waveform representation. A significant challenge is the non-uniformity of the sample data, which is a result of the mouse's energy-conserving behavior---the optical sensor does not report periodically if it is idle; instead, it will dynamically send packets as soon as the movement threshold is reached. For example, when the mouse is actively being used—for instance during typical office or home activities—the average polling rate will be close to the maximum for that mouse. For our target mice (See Table \ref{table:usedmice}), polling rate is either 4KHz or 8KHz. However, when the mouse is not in vigorous motion, the reported sample rate is significantly lower. This property is reflected in the collected data, as there are no reported samples where $\Vec{\Delta_P} = \Vec{0}$. 

To address the nonuniformity in the collected mouse signals, we apply a signal processing pipeline that includes sequential resampling and filtering stages.
We use a \verb|sinc|-based resampling technique that allows efficient and accurate interpolation and reconstruction of fragmented, nonuniform, and noisy signals \cite{eldarnonuniform}. 
This resampling stage is equivalent to applying a modified \textit{Whittaker–Shannon} interpolation formula:
\begin{equation}
    \hat{\Delta_P[t]} = \sum_{i = -\infty}^{\infty} \Vec{\Delta_P}[i] * sinc \left( \frac{t - t_i}{T} \right)
\end{equation}
For $T = \frac{1}{Fs}$ where $T$ and $Fs$ are the target period and the target polling rate respectively, for which we use $Fs = 16000$.

Lastly, we trim the generated signal to remove any silence/padding. After completing these steps, we save the mouse signal as a waveform, preparing it for the filtering stage in the Mic-E-Mouse pipeline.

\subsection{Waveform Signal Filtering}

It is important to note that environmental and sensor noise are a point of concern in our data collection scheme. Therefore, we implement a custom-tuned Wiener Filter adapted to the sensor noise introduced by having a high DPI setting. We collect 10-minute segments of the mouse jitter which are used to estimate the noise profile, and we couple the noise spectrum with the true spectrum of speech which we acquire from our selected datasets (see Sec~\ref{sec:datasets}). In real life scenarios, the attacker can obtain noise recordings by exploiting one of the methods discussed in Section \ref{sec:exploit} to collect data during times deep into the night when the user is expected to not be using the computer. If such information cannot be obtained, another alternative would be to tune the Wiener filter using the power spectral density of sections with the lowest overall signal energy as these are most likely the ones where the user is silent. By applying a Wiener filter, the attacker can efficiently increase the Signal-to-Noise Ratio (SNR) of the resulting waveform~\cite{wienerfilter, chenwiener}. 

\subsection{ML-based Signal Processing}
\label{sec:ML}
The last component of the Mic-E-Mouse pipeline uses ML-based models to extract speech features from the pre-processed mouse signal data and map them to an embedding space. These ML models are used for two main tasks: \textit{speech reconstruction} and \textit{speech keyword classification}. For each task, we provide in the following details about the models architectures and training parameters.

\subsubsection{Speech Reconstruction Models} 
This task involves reconstructing original speech signals from the recovered mouse data signals. We use a transformer encoder-only architecture inspired by the smallest configuration of the OpenAI Whisper model \cite{openaiwhisper}, which leverages multi-head residual self-attention blocks~\cite{attention, seq2seq}. We transform the combined $\Delta X$ and $\Delta Y$ data into a single 2-channel time-domain signal, then compute its log-mel-spectrogram (e.g., 80 Mel bins over 25 ms windows), forming the encoder input.
The model is trained on paired signals \((X, Y)\), where \(X\) denotes the noisy input (mouse data), and \(Y\) represents the high-quality ground-truth audio waveform~\cite{limspeech}. The training objective minimizes a loss function \(\mathcal{L}(Y, \hat{Y})\), where \(\hat{Y}\) is the reconstructed audio waveform. The loss is computed using L1 or L2 norm distance measures, defined as:
\begin{equation}
\mathcal{L}(Y, \hat{Y}) = \frac{1}{n} \sum_{i=1}^n \| Y_i - \hat{Y}_i \|_p    
\label{eq:loss_task1}
\end{equation}
where \(p = 1\) for L1 norm or \(p = 2\) for L2 norm. The model is optimized using the Adam optimizer with a learning rate of \(0.001\), training over 30 epochs. A step learning rate decay with \(\gamma = \frac{1}{\sqrt{2}}\) is applied every 5 epochs (equivalent to \(\gamma = 1/2\) at 10 epochs).
    
\subsubsection{Speech Keyword Classification Models}
This task involves the automated extraction of significant keywords from reconstructed audio waveforms. Formally, it is defined as a mapping function that translates the predicted audio waveform signal, denoted as $\hat{Y}$, into a set of key terms ${K_1, K_2, \ldots, K_n}$ that encapsulate critical information within the speech. We use wav2vec2\cite{wav2vec2}, designed to extract and classify keywords. Wav2vec2 is based on a transformer architecture, enabling a large temporal receptive field that spans 25 ms of audio. The Wav2vec2 model is trained on a supervised dataset, for 20 epochs each, with a batch size of 64 and Adam optimizer.

\subsubsection{Training Datasets}
\label{sec:datasets}
The ML models incorporated in the Mic-E-Mouse pipeline are trained using English voice data from existing speech recognition datasets, including (\emph{i}) the \textit{CSTR VCTK Corpus (version 0.92)} dataset, which contains approximately 54 hours of human speech recorded from 110 English speakers \cite{vctk}, and (\emph{ii}) the \textit{AudioMNIST} dataset, comprising 30,000 recordings of spoken digits in English collected from 60 speakers \cite{audiomnist}. The training data collection process involved playing each audio sample while simultaneously recording mouse sensor data (e.g., timestamps and movements). These sensor data were then aligned with their corresponding audio samples and ground-truth keywords.

\begin{figure}[ht!]
    \centering
    \includegraphics[width=.34\textwidth]{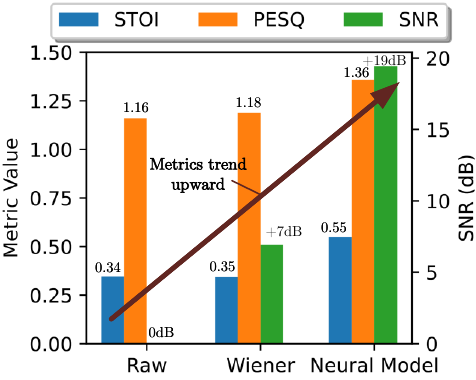}
    \caption{Waveform reconstruction metrics showing signal enhancement at each stage of the Mic-E-Mouse pipeline.}
    \label{fig:hist}
    \vspace{-0.4cm}
\end{figure}

\section{Evaluation}
\label{results}

\subsection{Evaluation Tasks and Metrics}
\label{subsec:tasks}

\noindent \textbf{Waveform Reconstruction.} 
We assess the efficacy of waveform reconstruction using the following metrics:
\begin{itemize}
    \item \textit{Short-Time Objective Intelligibility (STOI)}: 
    For measuring the intelligibility of reconstructed speech using a Discrete Cosine Transform~\cite{stoi}. 

    \item \textit{Perceptual Evaluation of Speech Quality (PESQ)}: For correlating the quality of reconstructed speech with estimated human perception~\cite{pesq}, 

    \item \textit{Scale-Invariant Signal-to-Noise Ratio (SNR)}: 
    For evaluating the quality of the reconstructed waveform \cite{snr}. 
\end{itemize}

\noindent \textbf{Keyword Classification.} 
To evaluate the effectiveness of keyword classification, we employ \textit{Accuracy}, which measures the ratio of correctly classified samples to their true classes.

\noindent \textbf{Human Assessment.} 
We focus on the following metrics:
\begin{itemize}
    \item \textit{Human Word Error Rate (HWER)}: 
    HWER divides the total number of errors in the reconstructed audio by the total number of keywords recognized by humans.
    
    \item \textit{Mean Opinion Score (MOS)}: 
    MOS evaluates the quality of audio signals as perceived by humans on a standardized scale ranging from 1 (poor) to 5 (excellent). 
    
    \item \textit{Semantic Similarity Ratings (SSR)}: 
    SSR evaluates the semantic similarity between the content of the reconstructed audio and the original audio. 
\end{itemize}


\section{Results}

\begin{figure*}
    \centering
    \includegraphics[width=.85\textwidth]{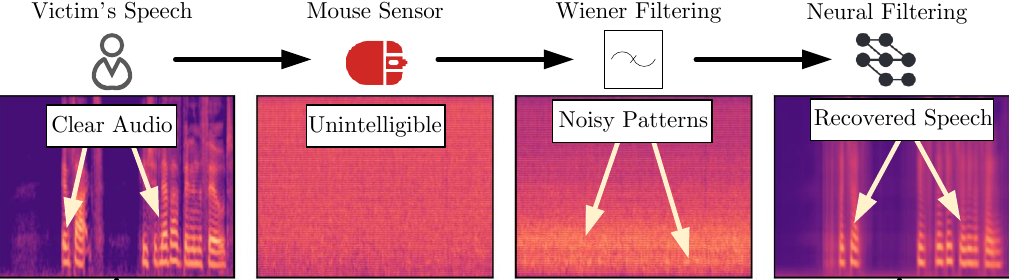}
    \caption{Reconstruction spectrograms showing the improvement following each stage.}
    \label{fig:spec}
    \vspace{-0.3cm}
\end{figure*}

\subsection{Attack Evaluation}

\subsubsection{Waveform Reconstruction}
Figures~\ref{fig:hist} and \ref{fig:spec} illustrate the results of speech audio reconstruction at various stages of the Mic-E-Mouse pipeline. The process begins with the raw mouse data signals, followed by Wiener filtering and preprocessing, and concludes with the final reconstruction using the OpenAI Whisper model, as detailed in Section~\ref{sec:ML}.

\begin{table}[ht!]
\centering
\caption{Classification accuracy across different tasks}
\scalebox{1.1}{
\begin{tabular}{l|cc} 
\toprule
Task          & \begin{tabular}[c]{@{}c@{}}Ground Truth \\Original Audio\end{tabular} & \begin{tabular}[c]{@{}c@{}}Mic-E-Mouse \\Reconstructed Audio\end{tabular}  \\ 
\midrule
MNIST Digit   & 99.1\%                                                                & 61.57 \%                                                                   \\
MNIST Speech & 92.8\%                                                                & 41.65 \%                                                                   \\
VCTK Speech  & 75.3\%                                                                & 62.30 \%                                                                   \\
\bottomrule
\end{tabular}
}
\label{table:classifiction}
\vspace{-0.2cm}
\end{table}

In Figure~\ref{fig:spec}, the original victim's speech audio waveforms are shown in the leftmost plot, depicting a clear and intelligible signal. The second plot shows the recorded mouse data signals when the same audio was played back, highlighting their high noise levels and unintelligibility. After applying Wiener filtering, the signals exhibit reduced noise and improved clarity, as shown in the next plot. Finally, the ML-based reconstruction recovers the original audio waveforms with high fidelity.

Figure~\ref{fig:hist} provides a quantitative evaluation of audio reconstruction performance using the three metrics discussed in Section~\ref{subsec:tasks}: STOI, PESQ, and SNR, computed across multiple instances of audio waveforms. The results indicate a significant improvement, with STOI increasing by 40\% after ML-based reconstruction compared to the raw data. Similarly, SNR shows a gain of +19 dB from an initial value of 0 dB. Additionally, the quality of the reconstructed audio improves progressively across the Mic-E-Mouse pipeline stages, as reflected by the linear increase in PESQ.

\subsubsection{Speech Classification}
We consider three main tasks: (\emph{i}) speech classification on the AudioMNIST dataset, (\emph{ii}) digit classification on the AudioMNIST dataset, and (\emph{iii}) speech classification on the VCTK dataset. The classification accuracy for each task is reported in Table~\ref{table:classifiction}, using the accuracy of the ground truth audio data as a reference. 
From Table~\ref{table:classifiction}, we observe that the classification accuracy on the reconstructed audio from the Mic-E-Mouse pipeline is approximately \(62\%\) for both the MNIST digit and VCTK speech recognition tasks. However, a lower accuracy of about \(42\%\) is observed for the AudioMNIST speech classification task, which can be attributed to the dataset size limitations of the AudioMNIST compared to VCTK.

\begin{table}[ht!]
\centering
\caption{Breakdown of human evaluation results.}

\resizebox{\linewidth}{!}{%
\begin{tabular}{l|ccc} 
\toprule
Method & HWER $(\downarrow)$ & MOS $(\uparrow)$ & SSR $(\uparrow)$ \\ 
\midrule
Raw & 100\% & 0 & 0 \\
Wiener Filtering & 100\% & 0 & 0 \\
Neural Model & \textbf{16.79\%} & \textbf{4.06} & \textbf{3.20} \\
Perceptual Neural Model & 44.92\% & 3.29 & 2.24 \\
\bottomrule
\end{tabular}
}
\label{table:userMOS}
\vspace{-0.2cm}
\end{table}

\subsubsection{Human Assessment}
We present the findings of a user study in which 16 human volunteers evaluated a selection of audio samples from the AudioMNIST dataset\footnote{This study received an exemption from an institutional IRB, and participants were guided through the process by team members.}. The study was conducted in a quiet office environment, where volunteers, informed about evaluating new "speech recovery systems," listened to 32 curated samples using Sony XM4 headsets. The relevant WER, MOS, and SSR metrics are summarized in Table~\ref{table:userMOS}. For the RAW and Wiener Filtering stages, all participants reported an absence of significant speech presence.

\begin{table}[ht!]
\centering
\caption{Classification accuracy across various audio volume levels (ranging from 50 dB to 80 dB)}
\begin{tabular}{l|cc} 
\toprule
Audio Volume & AudioMNIST Digit~ & AudioMNIST Speech~ \\ 
\midrule
80dB & 61.57\% & 47.65\% \\
70dB & 53.54\% & 32.58\% \\
60dB & 19.86\% & 21.28\% \\
50dB & 17.84\% & 16.05\% \\
\bottomrule
\end{tabular}
\label{table:volume}
\vspace{-0.3cm}
\end{table}

\subsection{Influence of environmental conditions}
\label{sec:conditions}
The accuracy of speech reconstruction from mouse signal data is significantly influenced by environmental factors, such as audio volume and surface material. We provide in the following a performance breakdown under different environmental conditions.

\subsubsection{Volume}
Table~\ref{table:volume} highlights the impact of varying volume levels on digit and speech classification accuracy. As shown, higher volume levels lead to improved accuracy, whereas lower volumes result in a sharp decline, underscoring the importance of adequate sound levels for reliable speech reconstruction using the proposed Mic-E-Mouse pipeline.

\begin{table}[ht!]
\centering
\caption{Speech recognition performance results across various surface materials (plastic, paper, and cardboard)}
\begin{tabular}{l|cc} 
\toprule
Material Surface & AudioMNIST Digit & AudioMNIST Speech \\ 
\midrule
Plastic & 61.57\% & 47.65\% \\
Paper & 55.20\% & 36.27\% \\
Cardboard & 23.06\% & 10.69\% \\
\bottomrule
\end{tabular}
\label{table:surface}
\vspace{-0.2cm}
\end{table}

\subsubsection{Surface Material}
Table~\ref{table:surface} demonstrates the impact of different surface materials on the classification accuracy of AudioMNIST Digit/Speech. The results show that smoother surfaces, such as plastic, achieve higher digit and speaker classification accuracies (61.57\% and 47.65\%, respectively) compared to rougher materials, like cardboard, which result in significantly lower accuracies (23.06\% and 10.69\%). These findings indicate that the type of surface beneath the mouse affects the clarity of the captured signal, thereby influencing the performance of the reconstruction process.

\subsection{Influence of mouse parameters}
Mouse parameters, such as polling rate and dots-per-inch (DPI), play an important role in the effectiveness of speech reconstruction. Table~\ref{tab:main-uni} presents the results of varying DPI settings across different polling rates. The results indicate that higher DPI settings generally yield better digit accuracy, particularly at higher polling rates like 8 kHz. Interestingly, the 2 kHz setting outperforms the 4 kHz setting, which may suggest suboptimal performance at 4 kHz due to hardware or signal processing limitations.

\begin{table}
\centering
\caption{AudioMNIST classification accuracy when using different mouse parameters (i.e., DPI and Polling rate)}
\resizebox{\linewidth}{!}{%
    \begin{tabular}{r|cc|cc|cc}
    \toprule
    & \multicolumn{6}{c}{Polling Rate}\\
    \cmidrule(lr){3-6}
    & \multicolumn{2}{c}{8kHz} & \multicolumn{2}{c}{4kHz} & \multicolumn{2}{c}{2kHz} \\
    \cmidrule(lr){2-3} \cmidrule(lr){4-5}\cmidrule(lr){6-7}
    DPI & Digit & Speaker & Digit & Speaker & Digit & Speaker \\
    \midrule
    20000 & 61.57\% & 47.65\%  & 38.45\% & 21.65\% & 52.29\% & 26.37\%\\
    15000 & 60.83\% & 26.96\%  & 34.29\% & 18.83\% & 43.65\% & 28.32\% \\
    10000 & 55.30\% & 22.74\%  & 27.49\% & 16.03\% & 32.78\% & 20.13\% \\
    \bottomrule
    \end{tabular}%
    }
    \label{tab:main-uni}
    \vspace{-0.3cm}
\end{table}

\subsection{Influence of signal degradation}

In addition to the inherent variability introduced by different mouse configurations, the Mic-E-Mouse pipeline can be further affected by signal degradation across three primary dimensions: (\emph{i}) timing jitter, (\emph{ii}) environmental noise, and (\emph{iii}) active mouse usage. Below, we detail the impact of each factor on speech recognition fidelity.

\subsubsection{Influence of timing jitter} 
\begin{figure}[ht!]
    \centering
    \includegraphics[width=.5\textwidth]{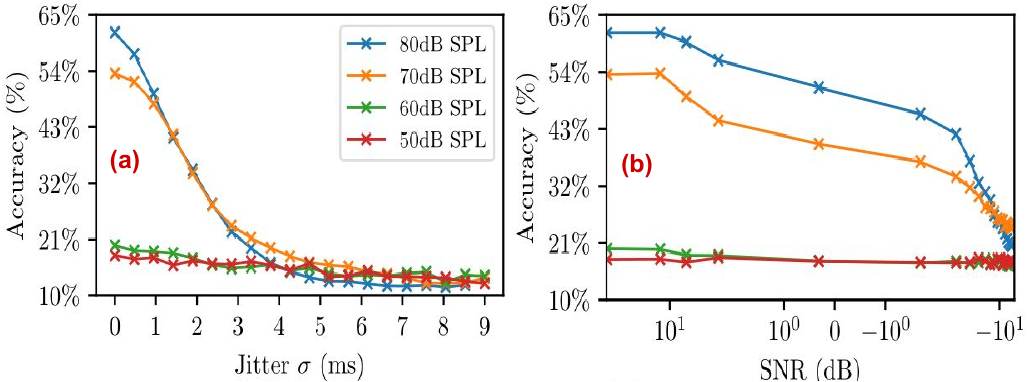}
    \caption{(\textbf{a}): Classification accuracy vs timing jitter and (\textbf{b}): Classification accuracy vs injected noise power (SNR).}
    \label{fig:jitter}
    \vspace{-0.3cm}
\end{figure}

Timing jitter refers to minute, random fluctuations in the intervals at which mouse data is reported or processed. Although each device advertises a nominal polling rate (e.g., 4,kHz or 8,kHz), real-world scheduling delays can cause the time between successive samples to deviate from the expected value. We posit that such jitter can distort the captured vibrations, particularly in higher-frequency ranges crucial for speech formats. We sample a timing jitter from a normal distribution $n = \mathcal{N}\left(0, \sigma \right)$ for each triplet $\left(\Delta T, \Delta X, \Delta Y\right)$, transforming it into $\left(\Delta T + n, \Delta X, \Delta Y\right)$, and we measure the accuracy of the trained models, we present the results in Figure~\ref{fig:jitter}(a).

\subsubsection{Influence of environment noise} While our controlled setup minimizes extraneous noise, real-world environments often include overlapping sounds from air-conditioning units, fans, traffic, or other nearby conversations. These unwanted vibrations can mask or interfere with the subtle signals the mouse sensor captures from speech. We sample two signal noises $\delta_X$ and $\delta_Y$ from a normal distribution $\delta = \mathcal{N}\left(0, \sigma \right)$ for each triplet $\left(\Delta_T, \Delta_X, \Delta_Y\right)$, transforming it into $\left(\Delta_T, \Delta_X + \delta_X, \Delta_Y + \delta_Y\right)$, and we measure the accuracy of the trained models, we present the results in Figure~\ref{fig:jitter}(b). 

\begin{table*}[t!]
\centering
\caption{Comparison of major side-channel eavesdropping works, highlighting the relation between channel capacity and the expected metrics.}
\resizebox{\linewidth}{!}{%
    \begin{tabular}{r|ccccccccccc}
    \toprule
    Related Work                & Sensing Modality  & Volume & Remote & Classification Model &Neural Filtering & Sampling Rate & Resolution & Channel Bitrate & SNR (dB) & Accuracy (\%) & STOI\\
    \midrule
    Lamphone~\cite{nassilamp}   & Optical       & 75-95dB   &  & - & - & 4kHz & 24 bits & 48 kbps & +24dB & - & 0.53\\
    Gyrophone~\cite{gyrophone}           & Mechanical    & 75dB      & \checkmark  & Dynamic Time Warping & - & ~200Hz & 16 bits & 3.2kbps &-&~26\%&-\\
    LidarPhone~\cite{LidarPhone}          & Laser         & 55-75dB   & \checkmark & CNN & - & 1.8kHz & 16 bits & 28.8 kbps &-&91\%&-\\
    Visual Microphone~\cite{visualmicrophone}   & Video         & 80-110dB  & & - & - & 2.2kHz & >10M bits & >22Mbps &+30dB&-&-\\
    mmEavesdropper~\cite{mmeavesdropper}      & Radar         & 60-90dB   & & LeNet-5& U-Net & - &-&-&+17dB& 93\%& -\\
    \midrule
    \rowcolor{gray!15}Ours      & Optical       & 50-80dB & \checkmark & Wave2Vec & Transformer & 8kHz  & 1.83 bits & 14.6 kbps & +19dB & 61\%& 0.55\\
    \bottomrule
    \end{tabular}%
    }%
    \label{tab:related_work}
    \vspace{-0.3cm}
\end{table*}

\section{Related Works}
\label{relworks}

This section outlines the similarities and differences between our approach and previous studies on sensor side-channel attacks. Table~\ref{tab:related_work} compares our proposed Mic-E-Mouse and related works. We observe that the experimental metrics across these studies are closely tied to both the \textbf{channel’s bitrate} and the intrinsic fidelity of its signal. For instance, the video-based side channel \cite{visualmicrophone}—operating at over 22~Mbps—delivers the highest SNR, whereas the optical channel of Lamphone \cite{nassilamp} with 48~kbps delivers worse SNR. The Gyrophone \cite{gyrophone} with its mechanical channel records the lowest accuracy owing to the low bitrate at just 3.2~kbps, and the use of Dynamic Time Warping (DTW) which performs worse than ML based approaches. Given its intermediate bitrate, Mic-E-Mouse naturally achieves mid-range performance in terms of the reported metrics

\subsection{Sensor-based Side-Channels}
Sensor-based Side-Channels are an ever-present threat in the modern digital landscape. Attacks on computer systems via compromised sensor and peripheral access have been shown to be effective at stealing privileged information~\cite{genkinaudio,farrukhpencil,liaomagear,yangyro,baaccel,chenmilli,humilliear,kwonghdd}. With the further integration of these technologies into consumer-facing products, the attack surface is poised to expand significantly. We note that motion sensors are a source of vulnerabilities as well, allowing for the reconstruction of stylus input through only the careful observation of sensor readings~\cite{farrukhpencil}. By exploiting access to sensor data in a similar way as the Mic-E-Mouse attack, the extraction of sensitive user data is feasible in much the same manner as these prior cyber-physical attack vectors.

Two works closely related to Mic-E-Mouse are LidarPhone and Lamphone. In LidarPhone, the authors demonstrate how the LiDAR sensor in robotic vacuum cleaners can detect minute surface deviations caused by air pressure fluctuations from speech. By interpreting the distance measurements from the LiDAR laser, they can reconstruct an audio signal that can be played back as needed. Similarly, Nassi et al. explored the use of light in Lamphone to detect tiny vibrations in a light bulb caused by sound waves. By using long-range optics, they measured these vibrations and filtered the sensor readings to recover the original sound that caused them.

Portable speakers and headphones are another weak point in the security model, as audio has been extracted using a magnetic side-channel attack~\cite{liaomagear}. This is done by targetting in-ear headphones and headsets, where the magnetic signals from the speaker diaphragm allow for information leakage. Notably, their method requires physical access to the victim and their vulnerable audio device. In the Mic-E-Mouse attack, we describe an attack that does not require physical proximity to the victim.

Cryptographic security is not immune to these types of attacks either~\cite{genkinaudio, liufreq, aydinseal, zhangvm, scheperswpa}. It has been shown that microphone signals can be used to decipher RSA keys without physical access to a victim's computer~\cite{genkinaudio}. This attack exploits the acoustic characteristics of computers in order to extract privileged data, much like the Mic-E-Mouse attack. However, the authors use a ground-truth audio signal to extract cryptographic keys, while we use collected mouse data to reconstruct a facsimile of a ground truth audio signal. 

In \cite{JerryAttack}, the authors presented an approach fundamentally different from our Mic-E-Mouse
\footnote{\cite{JerryAttack} was not published at the time of this paper’s submission,
our project \href{https://github.com/AICPS/Mic-E-Mouse}{GitHub Repo} started in May 2023, and \href{https://drive.google.com/drive/folders/1DcTldouupfp7BMteE1Br0lq7RCdQQ0Hc?usp=drive_link}{Data} was public in January 2024.
}
Their method relies on tampering with the mouse firmware to stream the raw (26x26) pixel image at 7-bit precision. These images, sampled at 3.7 kHz to reconstruct speech, require a bitrate of 17.5 Mbps, making the attack highly discoverable and of limited practical use. Furthermore, their threat model relies on an auxiliary microphone in the victim’s vicinity to perform full waveform reconstruction. Thus, Mic-E-Mouse is the first to demonstrate a practical and truly covert method, one that uniquely uses only standard user-space x-y coordinate data, which requires no physical or firmware modifications, and needs no external microphone.

\section{Discussion}
\label{discussion}

\subsection{Phoneme Intelligibility}
\begin{figure}[ht!]
    \centering
    \includegraphics[width=0.8\linewidth]{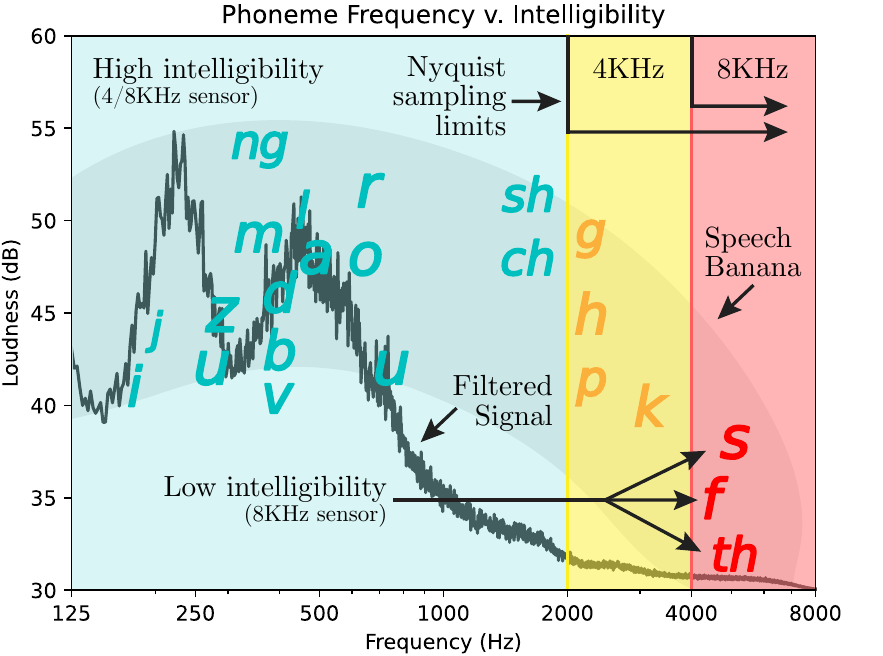}
    \caption{Phonemes primarily lie between 200Hz--2000Hz, shown here with an audio waveform overlay.}
    \label{fig:speech_banana}
    \vspace{-0.4cm}
\end{figure}

Figure~\ref{fig:speech_banana} illustrates the ``speech banana''~\cite{speechbanana}, covering typical speech frequencies (200Hz--2000Hz). Our overlay of the pipeline's filtered audio waveform confirms that this range corresponds to phoneme frequency and is thus captured reliably under the Nyquist–Shannon sampling theorem. Appendix~\ref{sec:speech_banana} provides additional details.

\subsection{Limitations}
\label{sec:limitations}

\noindent \textbf{Mouse Usage.} Frequent or concurrent mouse movements interfere with sensor inputs, degrading speech reconstruction. We assume limited mouse usage during sensitive speech (Section~\ref{threatmodel}). Additional hardware modifications (e.g., better motion tracking or noise-cancelation) could mitigate this.

\noindent \textbf{Surface Materials.} Only thin, flexible surfaces effectively transmit speech vibrations. Rigid or thick surfaces inhibit this side-channel, limiting broader applicability. Advances in sensor technology and ML-driven filtering may eventually relax these constraints.

\subsection{Countermeasures}
\label{sec:countermeasures}
\noindent \textbf{Mouse Pads.} Requiring a signal-absorbing mouse pad reduces vibration-based eavesdropping with minimal user disruption.

\noindent \textbf{Blacklisting Vulnerable Devices.} Devices featuring high-performance optical sensors (e.g., \rtextsc{Paw3395}, \rtextsc{Paw3399}) can be banned by IT policies. This can be automated at the OS level (e.g., via \verb|udev| rules~\cite{udevman}) to de-authorize high-risk USB HID devices.

\noindent \textbf{Approved Peripherals.} Institutions and remote workforces can maintain a curated list of safe mice to prevent the use of susceptible devices.

\section{Ethical Considerations}
\label{ethics}

Our research was conducted in a controlled laboratory with consenting participants. All collected speech data (both live and from existing datasets) belonged to the authors, and any code related to the Mic-E-Mouse attack remained in private repositories (i.e., not deployed to any primary open-source branch). We detail potential defenses (Section~\ref{sec:countermeasures}) to foster awareness of mouse-based eavesdropping risks.

\subsection{Responsible Vulnerability Disclosure}
\label{sec:vulndisc}
We have informed vendors of 26 affected products and the OEM sensor manufacturer (PixArt Imaging, Inc.) of these vulnerabilities. Mitigations typically require sensor or firmware updates that limit side-channel leakage while preserving user experience. We are also filing a CVE under ``CWE-1300: Improper Protection of Physical Side Channels,'' emphasizing the need for strengthened protocols in secure environments.

\section{Conclusion}
\label{conclusion}
The increasing precision of optical mouse sensors has enhanced user interface performance but also made them vulnerable to side-channel attacks exploiting their sensitivity. We demonstrate that an adversary can extract audio from a victim user using commercial mouse sensors (\rtextsc{Paw3395} and \rtextsc{Paw3399}), without physical access or a microphone. Our Mic-E-Mouse pipeline enables covert eavesdropping by leveraging non-uniform resampling, Wiener filtering, and transformer-based neural network filtering to enhance audio recovery. This multi-stage signal processing addresses challenges such as heavy quantization, non-uniform sampling, and high noise levels, producing comprehensible output signals. Our results highlight the feasibility and effectiveness of auditory surveillance via optical sensors.

\bibliographystyle{./IEEEtran}
\bibliography{ref}

\appendix

\subsection{The "Speech Banana"}
\label{sec:speech_banana}
We note that certain high-frequency phonemes--or perceptually distinct units of sound--such as \textit{f}, \textit{s}, and \textit{th} are significantly less distinguishable to our human evaluators. An explanation is derived from two important concepts -- (1) the Nyquist-Shannon sampling theorem \cite{shannonfs}, and (2) a mapping of common phonemes to loudness and frequency characteristics \cite{speechbanana}. In Figure \ref{fig:speech_banana}, we illustrate how the polling rate affects the intelligibility of selected high-frequency phonemes in the recovered audio waveform.

When a continuous-time signal is sampled at a rate less than double its highest frequency component, the resultant discrete waveform is not able to accurately represent the original continuous-time signal \cite{shannonfs}. For a sensor with a sample rate $F_s$ of 8KHz, we can obtain a useful discrete-time signal with maximum frequency up to $F_s/2$, or 4KHz. Similarly, for a 4KHz sensor, the maximum recoverable frequency then drops to 2KHz. Notably, the \textit{f}, \textit{s}, and \textit{th} phonemes have a characteristic frequency higher than even the Nyquist frequency for the 8KHz sensors. Moreover, these phonemes also have a below average loudness characteristic, further contributing to the low intelligibility observed by our human testers. 

\begin{figure}[!ht]
    \centering
    \includegraphics[width=\columnwidth]{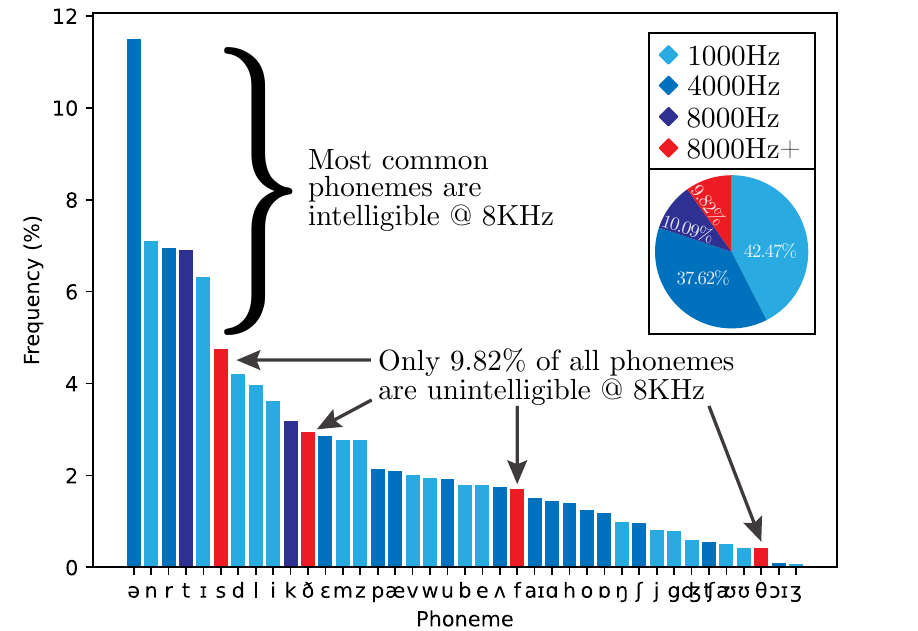}
    \caption{The distribution of phonemes and their associated minimum polling rate.}
    \label{fig:phoneme}
\end{figure}

We also note that the overwhelming majority (91.18\%) of phonemes in the English language are within the frequency range of the 8KHz sensor, and a significant portion (80.09\%) is still within the frequency range of the 4KHz sensor \cite{cmudict, phonemeanalysis, bnclist, advbiofreq, speechbanana}. These percentages are based solely from the theoretical lower bound on required sampling frequency obtained from the Nyquist-Shannon sampling theorem. In fact, we can see empirical evidence of this in Figure \ref{fig:speech_banana}, where the filtered signal strength drops off noticeably after the Nyquist frequency. Notably, even for a mouse with a 1KHz sensor, 42.47\% of phonemes are within this upper bound. The full distribution of phoneme prevalence versus required polling rate is shown in Figure \ref{fig:phoneme}.

\subsection{Additional Vulnerable Mice}
\label{sec:additionalmice}

\definecolor{mice_r}{HTML}{aa222d}
\definecolor{mice_o}{HTML}{cc3c00}
\definecolor{mice_y}{HTML}{cc8500}
\begin{table*}[!ht]\centering
\caption{An expanded selection of vulnerable mice (Red: DPI \& Poll Rate, Orange: Poll Rate only, Yellow: DPI only)}
\scalebox{1.0}{
\rotatebox[origin=c]{90}{increasing vulnerability}%
\hspace{0.5em}
\centering
\begin{tabular}{|l|c|c|c|c|c|c|}
    \hline
    \rtextsc{Vendor/Model} & \rtextsc{Sensor} & DPI & IPS& \rtextsc{Polling Rate (Hz)} & \rtextsc{Unit Price (USD)} & \rtextsc{Reconstruction}${}^*$\\
    \hline
    \hline
 	\textcolor{mice_r}{Razer Viper 8KHz } & \textcolor{mice_r}{\rtextsc{Paw3399} } & \textcolor{mice_r}{20,000 } & \textcolor{mice_r}{650 } & \textcolor{mice_r}{8,000 } & \textcolor{mice_r}{\$50} & \textcolor{mice_r}{91.18\%} \\
	\hline
	\textcolor{mice_r}{Darmoshark M3 } & \textcolor{mice_r}{\rtextsc{Paw3395} } & \textcolor{mice_r}{26,000 } & \textcolor{mice_r}{650 } & \textcolor{mice_r}{4,000 } & \textcolor{mice_r}{\$45} & \textcolor{mice_r}{80.09\%} \\
	\hline
	\textcolor{mice_r}{Darmoshark M34K } & \textcolor{mice_r}{\rtextsc{Paw3395} } & \textcolor{mice_r}{26,000 } & \textcolor{mice_r}{650 } & \textcolor{mice_r}{4,000 } & \textcolor{mice_r}{\$65} & \textcolor{mice_r}{80.09\%} \\
	\hline
	\textcolor{mice_r}{DeLUX M800 Ultra } & \textcolor{mice_r}{\rtextsc{Paw3395} } & \textcolor{mice_r}{26,000 } & \textcolor{mice_r}{650 } & \textcolor{mice_r}{4,000 } & \textcolor{mice_r}{\$60} & \textcolor{mice_r}{80.09\%} \\
	\hline
	\textcolor{mice_r}{Pulsar Gaming Gears X2H Mini } & \textcolor{mice_r}{\rtextsc{Paw3395} } & \textcolor{mice_r}{26,000 } & \textcolor{mice_r}{650 } & \textcolor{mice_r}{4,000 } & \textcolor{mice_r}{\$100} & \textcolor{mice_r}{80.09\%} \\
	\hline
	\textcolor{mice_r}{Rapoo VT3S } & \textcolor{mice_r}{\rtextsc{Paw3395} } & \textcolor{mice_r}{26,000 } & \textcolor{mice_r}{650 } & \textcolor{mice_r}{4,000 } & \textcolor{mice_r}{\$55}& \textcolor{mice_r}{80.09\%} \\
	\hline
	\textcolor{mice_r}{ThundeRobot ML901 } & \textcolor{mice_r}{\rtextsc{Paw3395} } & \textcolor{mice_r}{26,000 } & \textcolor{mice_r}{650 } & \textcolor{mice_r}{4,000 } & \textcolor{mice_r}{\$60} & \textcolor{mice_r}{80.09\%} \\
	\hline
	\textcolor{mice_r}{ThundeRobot ML903 } & \textcolor{mice_r}{\rtextsc{Paw3395} } & \textcolor{mice_r}{26,000 } & \textcolor{mice_r}{650 } & \textcolor{mice_r}{4,000 } & \textcolor{mice_r}{\$80} & \textcolor{mice_r}{80.09\%} \\
	\hline
	\textcolor{mice_r}{VGN Dragonfly F1 } & \textcolor{mice_r}{\rtextsc{Paw3395} } & \textcolor{mice_r}{26,000 } & \textcolor{mice_r}{650 } & \textcolor{mice_r}{4,000 } & \textcolor{mice_r}{\$35} & \textcolor{mice_r}{80.09\%} \\
	\hline
	\textcolor{mice_r}{Zaopin Z1 Pro } & \textcolor{mice_r}{\rtextsc{Paw3395} } & \textcolor{mice_r}{26,000 } & \textcolor{mice_r}{650 } & \textcolor{mice_r}{4,000 } & \textcolor{mice_r}{\$50}& \textcolor{mice_r}{80.09\%} \\
	\hline
	\textcolor{mice_r}{G-Wolves Hati S Plus } & \textcolor{mice_r}{\rtextsc{Paw3399} } & \textcolor{mice_r}{20,000 } & \textcolor{mice_r}{650 } & \textcolor{mice_r}{4,000 } & \textcolor{mice_r}{\$160}& \textcolor{mice_r}{80.09\%} \\
	\hline
	\textcolor{mice_o}{AtomPalm Hydrogen } & \textcolor{mice_o}{\rtextsc{Paw3360} } & \textcolor{mice_o}{12,000 } & \textcolor{mice_o}{250 } & \textcolor{mice_o}{8,000 } & \textcolor{mice_o}{\$100}& \textcolor{mice_o}{91.18\%} \\
	\hline
	\textcolor{mice_o}{AtomPalm Hydrogen 2 } & \textcolor{mice_o}{\rtextsc{Paw3360} } & \textcolor{mice_o}{12,000 } & \textcolor{mice_o}{250 } & \textcolor{mice_o}{8,000 } & \textcolor{mice_o}{\$100}& \textcolor{mice_o}{91.18\%} \\
	\hline
	\textcolor{mice_o}{Zaunkoenig M2K } & \textcolor{mice_o}{\rtextsc{Paw3360} } & \textcolor{mice_o}{12,000 } & \textcolor{mice_o}{250 } & \textcolor{mice_o}{8,000 } & \textcolor{mice_o}{\$350} & \textcolor{mice_o}{91.18\%} \\
    \hline
	\textcolor{mice_y}{Darmoshark N3 } & \textcolor{mice_y}{\rtextsc{Paw3395} } & \textcolor{mice_y}{26,000 } & \textcolor{mice_y}{650 } & \textcolor{mice_y}{1,000 } & \textcolor{mice_y}{\$40} & \textcolor{mice_y}{42.47\%} \\
	\hline
	\textcolor{mice_y}{DeLUX M800 Pro } & \textcolor{mice_y}{\rtextsc{Paw3395} } & \textcolor{mice_y}{26,000 } & \textcolor{mice_y}{650 } & \textcolor{mice_y}{1,000 } & \textcolor{mice_y}{\$50} & \textcolor{mice_y}{42.47\%} \\
	\hline
	\textcolor{mice_y}{Glorious Model O 2 } & \textcolor{mice_y}{\rtextsc{Paw3395} } & \textcolor{mice_y}{26,000 } & \textcolor{mice_y}{650 } & \textcolor{mice_y}{1,000 } & \textcolor{mice_y}{\$65} & \textcolor{mice_y}{42.47\%} \\
	\hline
	\textcolor{mice_y}{Keychron M3 } & \textcolor{mice_y}{\rtextsc{Paw3395} } & \textcolor{mice_y}{26,000 } & \textcolor{mice_y}{650 } & \textcolor{mice_y}{1,000 } & \textcolor{mice_y}{\$55} & \textcolor{mice_y}{42.47\%} \\
	\hline
	\textcolor{mice_y}{Redragon M994 } & \textcolor{mice_y}{\rtextsc{Paw3395} } & \textcolor{mice_y}{26,000 } & \textcolor{mice_y}{650 } & \textcolor{mice_y}{1,000 } & \textcolor{mice_y}{\$35} & \textcolor{mice_y}{42.47\%} \\
    \hline
	\textcolor{mice_y}{Edifier Hecate G3M Pro } & \textcolor{mice_y}{\rtextsc{Paw3399} } & \textcolor{mice_y}{20,000 } & \textcolor{mice_y}{650 } & \textcolor{mice_y}{1,000 } & \textcolor{mice_y}{\$70}& \textcolor{mice_y}{42.47\%} \\
	\hline
	\textcolor{mice_y}{Razer Basilisk V2 } & \textcolor{mice_y}{\rtextsc{Paw3399} } & \textcolor{mice_y}{20,000 } & \textcolor{mice_y}{650 } & \textcolor{mice_y}{1,000 } & \textcolor{mice_y}{\$40}& \textcolor{mice_y}{42.47\%} \\
	\hline
	\textcolor{mice_y}{Razer Basilisk V3 } & \textcolor{mice_y}{\rtextsc{Paw3399} } & \textcolor{mice_y}{20,000 } & \textcolor{mice_y}{650 } & \textcolor{mice_y}{1,000 } & \textcolor{mice_y}{\$80} & \textcolor{mice_y}{42.47\%} \\
	\hline
	\textcolor{mice_y}{Razer Basilisk Ultimate } & \textcolor{mice_y}{\rtextsc{Paw3399} } & \textcolor{mice_y}{20,000 } & \textcolor{mice_y}{650 } & \textcolor{mice_y}{1,000 } & \textcolor{mice_y}{\$105} & \textcolor{mice_y}{42.47\%} \\
	\hline
	\textcolor{mice_y}{Razer DeathAdder V2 } & \textcolor{mice_y}{\rtextsc{Paw3399} } & \textcolor{mice_y}{20,000 } & \textcolor{mice_y}{650 } & \textcolor{mice_y}{1,000 } & \textcolor{mice_y}{\$25}& \textcolor{mice_y}{42.47\%} \\
	\hline
	\textcolor{mice_y}{Razer DeathAdder V2 Pro } & \textcolor{mice_y}{\rtextsc{Paw3399} } & \textcolor{mice_y}{20,000 } & \textcolor{mice_y}{650 } & \textcolor{mice_y}{1,000 } & \textcolor{mice_y}{\$70}& \textcolor{mice_y}{42.47\%} \\
	\hline
	\textcolor{mice_y}{Razer Viper Ultimate } & \textcolor{mice_y}{\rtextsc{Paw3399} } & \textcolor{mice_y}{20,000 } & \textcolor{mice_y}{650 } & \textcolor{mice_y}{1,000 } & \textcolor{mice_y}{\$160} & \textcolor{mice_y}{42.47\%} \\
    \hline
    
\end{tabular}
}
\label{table:additionalmice}
\end{table*}

\begin{figure*}[!ht]
    \centering
    \includegraphics{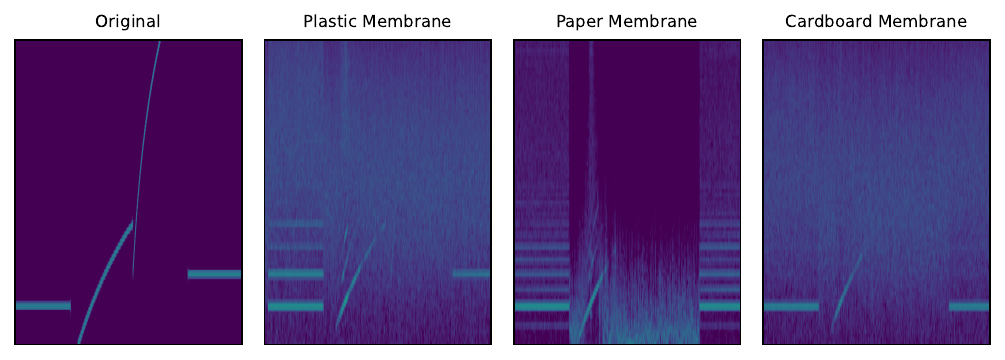}
    \caption{Comparison of the sweep response for the three selected membranes.}
    \label{fig:apx-membrane}
\end{figure*}

A list of similar mice that are also vulnerable to the same class of exploit as the Razer and Darmoshark is presented in Table \ref{table:additionalmice}. These additional mice share at least one of the following attributes: (1) a \rtextsc{Paw3395} or \rtextsc{Paw3399} sensor or (2) a polling rate at or above 4KHz.  We list sensors with a 4KHz or 8KHz. Table \ref{table:additionalmice} is ordered first by sensor vulnerability exposure and then alphabetically. Note that the asterisk in the last column of the table represents the theoretical upper bound on reconstruction percentage as shown in Appendix \ref{sec:speech_banana}. This reconstruction percentage is based on the polling rate of the sensor, whereas the overall vulnerability rating takes into account multiple other factors such as DPI, IPS, etc.

One important consideration is that certain mouse vendors market the sensors by using brand-specific names. An example of this is Razer using the brand name \textit{Focus+} in lieu of the actual sensor used-- he \rtextsc{Paw3399}. A similar case applies to Glorious Gaming's \rtextsc{BAMF2.0}, which is a derivative of the \rtextsc{Paw3395} sensor. We consider this kind of sensors as derivatives of the original PixArt Imaging sensors in our analysis and discussion. Table \ref{table:additionalmice} shows a list of 26 unique products using either the \rtextsc{Paw3395} or \rtextsc{Paw3399} sensors sorted according to their potential susceptibility to our attack.

We note the large size of the computer hardware market validates the growing ubiquity of these high-performance mice with state-of-the-art optical sensors. For example, Razer had total hardware sales of US\$1,452.4 million during Fiscal Year 2021 \cite{razer21report}, including desktops, laptops, and other computing equipment. 
For further evidence of the growing market demand, it's anticipated that the global market for computer mice will increase to US\$2.61 billion by the end of 2023, with a Compounded Annual Growth Rate (CAGR) of 8.1\%  \cite{mice23demand}. Then, it's clear that this type of high-performance mouse sensors will become more pervasive over time.

\subsection{Full Membrane Study}
\label{membrane}
We present a sweep response for various types of membranes, including paper, plastic, and cardboard. These are a representative sample of common types of surfaces a mouse may be placed on in a real-world environment. The sweep response, along with a comparison to the ground truth signal is presented in Figure \ref{fig:apx-membrane}.
The different responses are mainly attributed to three properties:
\begin{itemize}
    \item The thickness $W$ of the material significantly influences acoustic absorption and reflection. Thicker materials exhibit high sound absorption, leading to a dampened sweep response, whereas thinner materials may reflect sound more efficiently, resulting in an accentuated response.
    \item The density $\rho$ of the material dictates both the energy needed to vibrate the surface, as well as the momentum it carries before the signal dies down. Lighter materials, such as paper, are mismatched with the ambient noise and only acquire the injected sounds.
    \item The speed of sound in the material $\nu$. Materials with a higher $\nu$ facilitate rapid sound transmission, potentially yielding a less distorted response. In contrast, materials with a lower $\nu$ can delay wave propagation, altering the phase and amplitude response in the sweep.
\end{itemize}

Modeling the complete response given material properties is not trivial, given the existence of standing waves, non-linear responses, elastic attenuation, and potentially other physical phenomenon. However, we notice that these differences are mainly translated in two aspects in the extracted signals:
\begin{itemize}
    \item Harmonics: The emergence of 2nd and 3rd order harmonics in plastic and paper membranes suggests signal degradation. Nonetheless, this redundancy in frequency acquisition offers a compensatory mechanism, providing machine learning models with a 'second chance' to discern the underlying signal.
    \item The noise profile: The susceptibility to random noise varies significantly across materials. Notably, the paper membrane displays a distinct response, characterized by its 'quietness' in the sweep. This is attributed to a minimal presence of standing waves and rapid energy dissipation, owing to paper's low momentum energy, in contrast to cardboard and plastic.
\end{itemize}

\end{document}